\documentstyle[12pt,epsf]{article}

\begin{document}
\begin{flushright}
{\bf JYFL 97-11}\\
{\bf{FIAN/TD-11/97}}
\end{flushright}
\bigskip
\bigskip
\begin{center}
{\bf{Dilepton Emission from Dense Hadron Gas}}
\end{center}
\bigskip
\begin{center}
{\bf A.V. Leonidov$^{(1)}$ and P.V. Ruuskanen$^{(2)}$}
\end{center}
\bigskip
\begin{center}
{\it{ (1) P.N.~Lebedev Physics Institute, Moscow, Russia}}
\end{center}
\begin{center}
{\it{ (2) University of Jyv{\"a}skyl{\"a}, Jyv{\"a}skyl{\"a}, Finland}}
\end{center}
\bigskip
\begin{center}
{\bf Abstract}
\end{center}
Using a Hagedorn resonance gas picture and quark-hadron
duality we estimate the dilepton emission rate in the vicinity of
the QCD deconfinement phase transition. The result is then used to
calculate a dilepton spectrum in ultrarelativistic heavy ion
collisions. We show that multibody contributions taken into account in
the Hagedorn resonance gas approach  provide an enhancement of
the production rate of massive dileptons
as compared to the previously considered sources.

\newpage

\section{Thermal rate from a dense hadron matter}

  The dilepton production in high energy heavy ion collisions
has drawn considerable attention as a possible signal for the appearance
of a quark-gluon plasma \cite{or}, see also the reviews \cite{rev}.
The main problem, however, is to
distinguish the signals coming from the normal hadronic phase and
from the quark-gluon plasma.
To be able to do so one needs reliable models for the dilepton production.
At extremely high temperatures the situation is relatively clear. Due
to the asymptotic freedom of QCD the dominant production mechanisms are
perturbative.
At very low temperatures we also have a reliable
description of dilepton production in a rarefied hadron gas,
predominantly by pion annihilation.
Much less can be said about the
intermediate temperature range around the phase transition temperature
$T_c \sim 160$ MeV, where the hadronic gas is very hot and dense and the
usual methods do not work,
but the quark-gluon plasma is cold and rarefied
and the perturbative QCD results are less reliable.
We shall address the problem of estimating the dilepton emission from a
hadron gas at these intermediate temperatures.
 This is of special interest because
a hadron gas close to the transition temperature $T_c$ could be a dominant
feature of the final state produced at the present CERN heavy ion experiments
with lead beams.

The new experimental data on dilepton production in heavy ion collisions
demonstrate a pronounced excess of produced dileptons over those produced by
known hadron sources both in low-mass range \cite{CERES} and
at larger masses \cite{NA50}.
 This is a clear
motivation to look for additional sources of dileptons not taken
into account in the conventional treatment.
The excess of small-mass dileptons has been extensively
discussed (see, e.g., \cite{BR,CERESexpl}).
More related to our discussion is, however, the excess
in the intermediate mass range 1.5 GeV $\le M \le$ 2.5 GeV by roughly
a factor of 2 with respect to expected contributions from the known
sources. This is
because the theoretical description we are developing is valid only for
sufficiently large invariant masses of lepton pairs. The results will be
shown for the mass interval 1.5 GeV $\le M \le$ 4 GeV.
We shall argue, that by improving the treatment of dilepton emission from
the dense and hot hadron gas at the vicinity of $T_c$
one can get such an increase with respect
to the conventional description.

 In \cite{LR} we have proposed to use a Hagedorn resonance gas
picture and quark-hadron duality to calculate the dilepton emission rate in
the hadron phase at temperatures close to the critical one. Below we shall
refine these arguments and present additional support to the main conclusion
of \cite{LR}. In the next section we shall construct an interpolation for
the dilepton emission rate satisfying the conditions following from the
quark - hadron duality and use it in a hydrodynamical model for heavy
ion collisions. In the last section we formulate our conclusions.

The problem of calculating the dilepton emission rate in a hadron gas
including also the mass range 1.5 GeV $\le M \le$ 4 GeV has been
considered in \cite{Lichard} and \cite{Huang}. We compare
the results of \cite{Lichard,Lichard2} for the rate from binary
hadron collisions
with our results from the dense resonance gas at $T_c$.
We shall comment on the relation of our results to those of \cite{Huang}.
Since our treatment effectively averages over mass intervals of the order of
vector meson widths and mass differences, we cannot address the problem of
medium effects on individual resonance parameters, relevant for the low-mass
region of CERES results \cite{CERES}.

 Let us first remind of the picture of  dilepton
production at a critical temperature proposed in \cite{LR}.
The main idea can be formulated as follows. In the vicinity
of the critical point the hadronic matter is extremely dense
B
and the interactions between the hadrons are very strong.
The idea of a Hagedorn resonance gas approach \cite{Hag} is
to encode the infinitely complicated strong interaction
physics of a hot and dense hadron matter in a single function,
an exponentially rising mass spectrum
\begin{equation}
\rho (m) = c {1 \over m^a} e^{m/T_0}\,.
\label{spectrum}
\end{equation}
The main idea proposed in \cite{LR} was that the exponential
spectrum in Eq.~(1) takes care of strong interactions only. The
free massive degrees of freedom saturating the partition function
of the Hagedorn gas can still experience electromagnetic and weak decays.
Of a particular interest to us is the spectrum of vector mesons, which
are the only important source of dilepton production in the Hagedorn resonance
gas in the considered mass range.

In the vector dominance model, to be used below, the dilepton width of a
vector meson is given by
\begin{equation}
\Gamma^{l+l-}_M = {1 \over g^{2}(M)} {4 \alpha^2 M \over 3}\,.
\label{width}
\end{equation}
The spectrum of dileptons produced by the decays of vector resonances
belonging to the Hagedorn gas can be written in the
relativistic kinetic theory in the form
$$
{dR^{l+l-}_V \over dM^2} =
\int dm \rho_V (m) \int {d^3 p \over 2E (2 \pi)^3}\ {\rm{exp}}(-E/T)
\int {d^3 p_1 \over 2E_1 (2 \pi)^3} {d^3 p_2 \over 2E_2 (2 \pi)^3}
$$
\begin{equation}
\times {\overline{|{\cal{M}}(p \to p_1+p_2)|^2}}\,
 (2 \pi)^4 \delta^{(4)} (p-p_1-p_2)
\delta((p_1+p_2)^2 - M^2)\,,
\label{kinrate}
\end{equation}
where $p$ is the four-momentum of a decaying vector meson with a mass $m$,
$E=\sqrt{{\bf{p}}^2+m^2}$ is its energy,
$p_{1,2}$ are the four-momenta of the emitted lepton and antilepton, and
$\rho_V (M)$ is the subspectrum of vector mesons in Eq.~(\ref{spectrum}),
not necesserily exponential. In Eq.~(\ref{kinrate}) we have assumed
a Boltzmann distribution for the energy of the decaying meson which holds
with a very good accuracy provided the invariant mass is high enough.

In the above formula the squared matrix element for the $V \to l^+ l^-$
decay, including spin summation and averaging, has the form
$$
{\overline{|{\cal{M}}(p \to p_1 + p_2)|^2}} =
  \left({4 \pi \alpha \over g(m)}\right)^2\, {4 \over 3}\, (p_1 + p_2)^2\,,
$$
which also gives the the result Eq.~(\ref{width}) for the decay width.
For the dilepton emission rate one gets from Eq.~(\ref{kinrate})
\begin{equation}
{dR^{l+l-}_V \over dM^2} =
 \rho_V (M)\ {\alpha^2 \over 6 \pi g^2 (M)}\ M^2\, T K_1 (M/T)\,.
\label{resrate}
\end{equation}

The most important point
is now to fix the mass dependence of the coupling constant $1/g(M)$.
Here we can use the known cross section of $e^+ e^-$
annihilation into hadrons which takes place in the same vector channel we
are considering. The cross
section of $e^+e^-$ annihilation into heavy vector mesons characterized by
the mass spectrum $\rho_V (M)$ reads:
\begin{equation}
\sigma (e^+e^- \rightarrow V) =
{(2 \pi)^3 \alpha^2 \over g^2(Q)}\, {1 \over Q}\, \rho_V(Q)\,,
\label{annres}
\end{equation}
where $Q$, the CMS energy of $e^+e^-$ collision, equals
the mass of produced vector meson.
We proceed by assuming in the spirit of the Hagedorn bootstrap picture
that the total cross section of $e^+ e^-$
annihilation into hadrons is saturated by the production of vector meson
states. This allows us to relate the emission rate from
the Hagedorn gas to the total $e^+e^-\to {\rm hadrons}$ cross section.
We first have
\begin{equation}
\sigma(e^+e^- \to {\rm hadrons}) =
R^{exp}(Q) {4 \pi \over 3} {\alpha^2 \over Q^2} =
\sigma(e^+e^- \to V) (Q)\,.
\label{dual}
\end{equation}
From the above two  equations we get the
the mass dependence of the coupling constant
\begin{equation}
{1 \over g^2 (M)} = {R^{exp} (M) \over 2 \pi^2} {1 \over M}
 {1 \over \rho_V (M)}
\label{coupling}
\end{equation}
and finally, by inserting this into Eq.~(\ref{resrate}),
we can calculate the dilepton production rate in terms of known quantities:
\begin{equation}
{dR^{l+l-}_V \over dM^2 } =
 R^{exp}(M) {\alpha^2 \over 6 \pi^3 } M T K_1 (M/T)\,.
\label{resrate1}
\end{equation}
Let us rewrite the function $R^{exp} (M)$ as
\begin{equation}
R^{exp} (M) \equiv R^{part} + \delta R (M)\,,
\label{Rdec}
\end{equation}
where we have decomposed $R^{exp} (M)$ into a sum of its parton model
prediction
$R^{part} = N_c \sum_f e^2_f$,
where $e_f$ is an electric charge of the quark in units of the electron
charge, and the deviation from the parton model result, $\delta R (M)$.
Experimentally one has $R^{exp} \sim R^{part}$ with a good accuracy
 for the mass-averaged cross section even though
the local mass spectrum shows a visible resonance structure
parametrized by $\delta R$.
From Eqs.~(\ref{resrate1},\ref{Rdec})
we get
$$
{dR^{l+l-}_V \over dM^2 } =
 R^{part} {\alpha^2 \over 6 \pi^3 } M T K_1 (M/T) +
 \delta R (M) {\alpha^2 \over 6 \pi^3 } M T K_1 (M/T),
$$
or finally
\begin{equation}
{dR^{l+l-}_V \over dM^2 } = {dR^{l+l-}_q \over dM^2} + O(\delta R)\,.
\label{resratef}
\end{equation}
This result shows that the parton model emission rate of
dileptons with mass $M$ equals that from the resonance gas at $T \sim T_c$
with the
same accuracy as parton model describes the total cross section for $e^+e^-
\to {\rm{hadrons}}$ at the CMS energy $M$. We should like to stress, that
this result does not depend on a particular assumption on the
form of the heavy vector meson spectrum.
A natural interpretation of the result is
that hadrons as local annihilating degrees of freedom
do not provide a simple description of the
dilepton production  in the
immediate vicinity of a critical point, where they start to dissolve
to their partonic constituents. As the correlations among the constituents get
weaker the electromagnetic current, when probed through the production of
{\it heavy} lepton pairs, is effectively that of quarks.

This picture leads us in turn to a conjecture on the general behaviour of the
dilepton rate as a function of temperature. Let us recall, that a formal
nonperturbative expression for the dilepton rate in the hadron gas reads
 \cite{MT}, \cite{W}
\begin{equation}
{d N \over d^4 q} = - {\alpha^2 \over 6 \pi^3 q^2} \int d^4 x
e^{-iqx} << J^{\mu} (x) J_{\mu} (0)>>_T.
\label{nonp}
\end{equation}
Basing on Eq.~(\ref{resratef}) and the above-described physical picture
it is tempting to
suggest that at high enough dilepton masses the current-current correlator
in (\ref{nonp}) and thus the dilepton emission rate is to a good accuracy a
continuous function of temperature across the phase transition point. How low
in the dilepton invariant mass does this hold is determined by the accuracy
related to the approximation made in Eq.~(\ref{dual}).
It would be interesting to check this conjecture in lattice stimulations.

As our aim is to calculate dilepton production in heavy ion collisions, we
shall need a continuous rate interpolation from the hadron rate to the QCD one
as a function of temperature.  A simple way of constructing an interpolation
which will agree with Eq.~(\ref{resratef}) is to start from
the rate due to binary hadron collisions
calculated in \cite{Lichard} and match it to the QCD one at the
critical temperature $T_c$.

Before turning to the actual hydrodynamic calculation, let us
also mention a calculation in \cite{Huang} based on using the
soft pion theorems in calculating the hadronic contribution to the
polarization tensor. The main result of this paper for thermal dilepton
rate from the hadron gas can be rewritten in the form
\begin{equation}
{dR^{l+l-}_{(s)h} \over dM^2} = {dR^{l+l-}_q \over dM^2}
\left[1-(\varepsilon- {\varepsilon^2 \over 2})
({\rho^V(M) - \rho^A(M) \over \rho^{em}(M)})\right]
(1+\delta R / R^{part})\,,
\label{specrate}
\end{equation}
where $\varepsilon = T^2/6F^2_{\pi}$, $F_{\pi}=93$ MeV and
$\rho^{em}$, $\rho^V$ and $\rho^A$ are the suitably defined electromagnetic,
vector and axial spectral densities which can be extracted from measured
quantities (for details see \cite{Huang}).
In the following we shall be interested in
the production of dilepton pairs having invariant masses $M > 1.5$ GeV.
In this mass range one can neglect $\rho^A$ with respect to $\rho^V$ and
take $\rho^V = \rho^{em}$ \cite{Huang}.
 This gives, when neglecting $\delta R$,
\begin{equation}
{dR^{l+l-}_{(s)h} \over dM^2} = {dR^{l+l-}_q \over dM^2}
[1-(\varepsilon- {\varepsilon^2 \over 2})]\,.
\label{specrate1}
\end{equation}
It is interesting that the temperature dependence through $\varepsilon$
is in fact quite smooth,
so the rate is predicted to be close to the thermal
quark-antiquark one even at temperatures below $T_c$.
This obviously agrees with
our main result Eq.~(\ref{resratef}). Unfortunately the range of masses
where experimental information is available for the technique of \cite{Huang}
is limited to $M \le 2.2$ GeV, so to analyze a broader mass interval
one in forced to use a more phenomenological approach as described below.

\begin{figure}
\epsfysize=8.5cm
\centerline{\epsffile{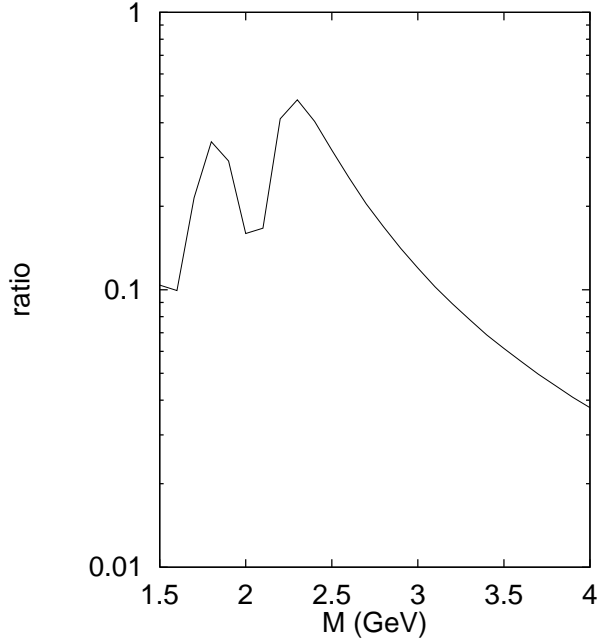}}
\caption{Ratio of static hadronic dilepton production rate in a binary
collision approximation to the quark-antiquark one at T=160 MeV}
\label{fig1}
\end{figure}

Let us now turn to the calculation of the dilepton emission
rate from a hadron gas due to binary collisions of mesons \cite{Lichard}.
In view of the previous discussion it is of special interest and importance
to look how close is the rate of this approach to our result.
This will provide valuable information on the relative importance
of the higher resonances and the
multibody effects in dilepton emission processes. In the Boltzmann
approximation for the thermal distribution, which works fairly well when
the invariant masses of lepton pairs satisfy $M > 1.5$ GeV,
the rate reads
\begin{equation}
{dR^{l+l-}_{(b)h} \over dM^2 } =
 \alpha^2\, {\sigma_h (M) \over (2\pi)^4}\, M T K_1(M/T)\,,
\label{hrate}
\end{equation}
where $\sigma_h (M)$ is calculated from the specific Lagrangian for meson
interactions \cite{Lichard}. In the following we shall use the rate
Eq.~(\ref{hrate}) including the following reactions \footnote{We are very grateful
to P.~Lichard for providing us the numerical data for this rate.} (for brevity
we list only the incoming particles) \cite{Lichard2}:
 $\pi^+ \pi^-$, $ K^+ K^-$, $K^0 {\bar K}^0$,
$\rho^+ \rho^-$, $K^{*+} K^{*-}$, $K^{*0} {\bar K}^{*0}$, $\pi \rho$,
 $\pi^0 \omega$, $\pi^0 \phi$, $\eta \rho^0$, $\eta \omega$, $\eta \phi$,
 $\eta' \rho^0$, $\eta' \omega$, $\eta'  \phi$, $K^+K^{*-} + K^-K^{*+}$,
 $K^0 {\bar K}^{*0}+{\bar K}^0 K^{*0}$, $\pi^0 \omega(1420)$,
 $\pi^0 \omega(1660)$. In Fig.~\ref{fig1} we show the ratio of this binary
hadron rate to our result Eqs.~(\ref{resrate1}) and (\ref{resratef})
at $T = 160$ MeV in the mass interval of interest.
%-------------------------------------------------------------Figure1
%------------------------------------------------------------------------
We see that the rate from binary interactions is less than our rate which
takes into account
also higher resonances and multibody effects
and, in the approximation we use, equals the rate from quark-antiquark
annihilations at this temperature.
In the Hagedorn resonance gas approach the critical temperature is
expected to be of order ($T_c \sim 160$ MeV), so taking into account
contributions
from binary collisions only is not sufficient at least in the vicinity of the
critical point. In the end of the next section we shall discuss the numerical
importance of this in calculating dilepton production in heavy ion collision
in a hydrodynamic model, but already now one could notice, that the
dependence on temperature in the above ratio cancels
(cf.~Eqs.~(\ref{resrate1}) and (\ref{hrate}), so the
difference is directly translated into the changes in the total dilepton rate.

\section{Dilepton emission in a hydrodynamic model}

In this section we shall study the implications of the above considerations
on dilepton production in heavy ion collisions described with the simplest
one-dimensional scaling hydrodynamic model \cite{Bjorken}. The resulting
dilepton rate is obtained by integrating over the space-time history
of the expansion.

The main result of the previous section is the approximate continuity of the
thermal emission rate of massive lepton pairs across the phase transition
point.
Thus the expression for the thermal dilepton emission in hadron phase used in
hydrodynamical calculations should be constructed to ensure this continuity
matching. In the narrow mass range, where the experimental data on spectral
densities is available one can apply Eq.~(\ref{specrate}) with the
approximate continuity automatically satisfied.
To describe the dilepton
mass spectrum in a wider mass range, as needed to compare with experiment,
one needs effective approaches like that in
\cite{Lichard}. Since the binary reactions should be the main source of
dileptons at low densities and temperatures and since we expect our results to
apply best in the vicinity of the critical temperature, we will
construct an interpolation procedure which at $T_c$ equals the quark matter
rate and goes over to the binary hadron rate of \cite{Lichard} a low
temperatures.

Below we shall consider the cases of first order and second order
deconfinement phase transition.
 The critical temperature in both cases is chosen
to be $160$ MeV and the freezeout temperature $130$ MeV.
The basic difference between these two cases is the presence or
absence of a mixed phase. From the arguments of the previous
section it is clear, that for the dilepton emission the rate should stay
equal to the quark-antiquark one for the whole duration of mixed phase.

Let us rewrite a well-known basic formula for the spectrum of dileptons
emitted during the one-dimensional scaling expansion (see e.g. \cite{KM},
\cite{Ru}) in the form
\begin{equation}
{dN^{l+l-} \over dM dy} = \pi R_A^2\
\int_{\tau_1}^{\tau_2} d \tau \tau\ r^{q(m,h)}_{q(h)}(T(\tau))
{dR^{l+l-}_{q(h)} \over dM^2}(M,T(\tau))\,,
\label{hydro}
\end{equation}
where $R_A$ is the radius of the smaller of the two colliding nuclei,
$r^{q(m,h)}_{q,h}$ specifies the weight for the
quark matter and hadron gas thermal rates (lower index)
in each of the phases, quark matter, mixed
or hadron gas (upper index). A head-on collision is assumed.
To specify completely the evolution of the system one has to
know the equations of state for both phases and the nature of
the  phase transition.
The entropy of the quark-gluon phase will be taken equal to that of the
ideal gas of gluons and three lightest quarks
\begin{equation}
S_q(T) = {74 \pi^2 \over 45} T^3+{1 \over \pi^2}
(12 m_s^2 T K_2(m_s/T)+3 m_s^3 K_1(m_s/T))\,.
\label{qen}
\end{equation}
A hadron phase will be described by a free resonance gas including all
particles with the masses up to 2 GeV. The dependence of temperature on
proper time $T(\tau)$ follows from the Bjorken equation \cite{Bjorken}
\begin{equation}
S(T(\tau)) = {1 \over \tau} {3.6 \over \pi R^2_A} {dN \over dy}\,.
\end{equation}

 Let us start with the case when the deconfinement transition from
the hadron gas to the quark-gluon plasma is of second order. Here the
entropy is continuous across the phase transition point $T_c$. We shall
assume that at all temperatures the entropy is given by a slightly
generalized form of an interpolation used in~\cite{Blaizot})
\begin{equation}
S(T) = (1-\eta (T)) S_h (T) + \eta (T) S_q (T)\,,
\label{entropy}
\end{equation}
where
\begin{equation}
\eta (T) = \frac{1}{2} (1+{\rm{tanh}}[(T-T_c)/\Delta T])
\label{eta}
\end{equation}
and $\Delta T$ specifies the width of the transient zone where the
entropy density of one phase goes over to that of the other phase.
We shall take $\Delta T = 16$ MeV, 10 \% of $T_c$. Let
us note, that the interpolation (\ref{entropy}) does not imply that the
phases are mixed at any temperature, it just provides a smooth
interpolation between the two known asymptotic expressions.
The dilepton emission rate is equal to that from quark matter,
$dR^{l+l-}_q / dM^2$, above $T_c$,  i.e. the corresponding weights in
Eq.~(\ref{hydro}) at $T > T_c$ are $r^q_h = 0$ and $r^q_q = 1$.
Below $T_c$ it is chosen to be a combination of the quark matter and
hadronic rates:
\begin{equation}
\left.{dR^{l+l-} \over dM^2}\right|_{T<T_c} =
 r^h_q (T) {dR^{l+l-}_q \over dM^2} + r^h_h (T) {dR^{l+l-}_{h}
\over dM^2}\,.
\label{rateint}
\end{equation}
The weights in this equation are
\begin{equation}
\left\{
\begin{array}{rcl}
r^h_q (T)& = & \eta (T)/(1 - \eta (T)), \\
r^h_h (T) & = & 1-r_q (T) \\
\end{array}
\right.
\label{int}
\end{equation}
ensuring that at $T = T_c$ the overall rate equals that from
the quark matter and providing an interpolation between
the quark rate at $T_c$ and hadron rate at low temperatures.
 As discussed in detailed in the previous
section, the necessity of an extra contribution
in the hadronic phase is due to the necessity
 of taking into
account the effects beyond the binary collision approximation, in which
the results of \cite{Lichard} were obtained. Otherwise
Eq.~(\ref{resratef}) will not hold at $T=T_c$.

\begin{figure}
\epsfysize=8.5cm
\centerline{\epsffile{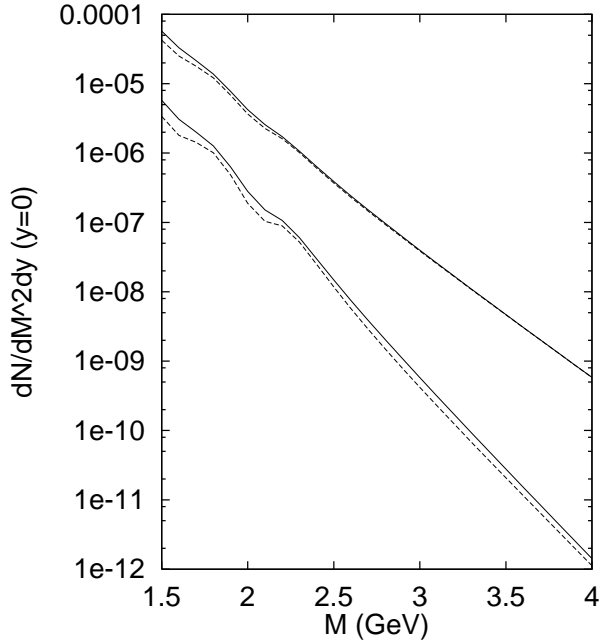}}
\caption{Dilepton spectra for SPS (lower curves) and RHIC (upper curves)
for second order phase transition. The dashed curves correspond to a binary
collision approximation to the hadronic dilepton rate, the solid ones
correspond to the rate interpolation Eq.~(\ref{rateint}).}
\label{fig2}
\end{figure}

In Fig.~\ref{fig2} we show the dilepton spectra calculated for two
different particle densities corresponding to head-on Pb-Pb collisions
at zero rapidity at SPS
($dN/dy=600$, lower curves) and RHIC ($dN/dy = 1500$, upper curves)
energies.  The dashed curves correspond to the binary collision
approximation to the hadronic dilepton rate and the solid ones
to the rate interpolation Eq.~(\ref{rateint}).
%-------------------------------------------------------------Figure2
%------------------------------------------------------------------------
We see that at SPS energies the contribution from the extra sources in hadron
phase which are not taken into account in the binary collision approximation
with a limited number of interacting mesons
is visible in the resulting dilepton spectrum and brings a multiplicative
factor of the order of $1.5$.  At RHIC energies the signal is already almost
totally dominated by pairs from quark-antiquark annihilations at the hot early
stages of the collision ($T \ge T_c$). Thus the details of the description of
the hadronic electromagnetic currents in this mass range, even in the vicinity
of $T_c$, are no longer important.

\begin{figure}
\epsfysize=8.5cm
\centerline{\epsffile{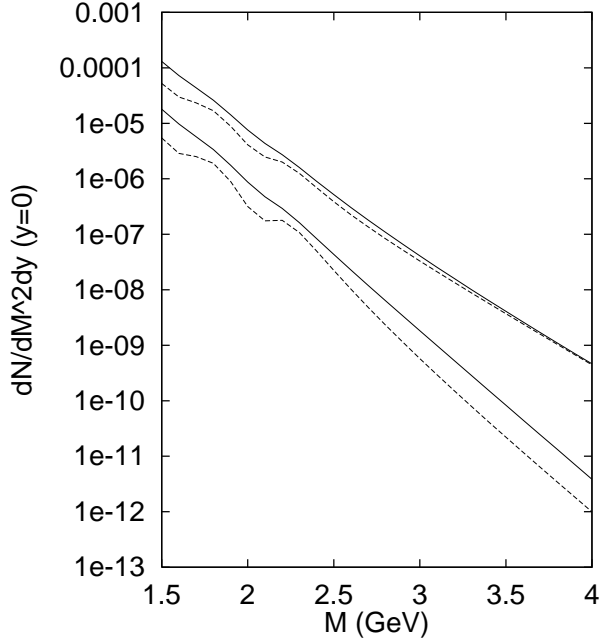}}
\caption{Dilepton spectra for SPS (dashed line) and  RHIC (solid line)
 for the first order phase transition.}
\label{fig3}
\end{figure}

Let us now consider the case of a first order phase transition. The
basic difference from the previous case is that the matter evolves
from the quark phase to the hadron one through a mixed phase.
Since in our approach at $T=T_c$ the rate in hadron phase approximately equals
that in quark phase, the weights in the mixed phase are $r^m_q=1$ and
$r^m_h=0$. This enhances the contribution from the
mixed phase as compared to the standard scenario
in which only binary hadron collisions are producing dileptons from
a volume occupied by hadronic phase throughout the time the mixed phase
exists (cf.~Fig.~1).
In the hadronic phase, $T < T_c$, the rate is taken
to be the same superposition of quark and hadron rates,
Eqs.~(\ref{rateint}) and (\ref{int}), as in the case of second order
transition. The results of calculation for Pb-Pb collisions
at SPS and RHIC energies are plotted
in Fig.~\ref{fig3}.
%-------------------------------------------------------------Figure3
%------------------------------------------------------------------------
Here the picture is somewhat more complicated than in the case of second
order transition. The main conclusions can be formulated as follows. At SPS
energies
the system is created at a temperature close to a critical one, so the main
contribution is coming from the mixed and low-temperature phases. The
new sources enhance the dilepton rate approximately with a factor of
3 - 4 in the mass range under consideration. The contributions
originating
from the new sources in the mixed and the low-temperature phase are of
the same order. At RHIC energies the contribution of the high temperature
phase is already quite significant at $M=2$ GeV and dominates at $M=4$ GeV.
The new sources enhance the rate with a factor 2 at $M \sim 2$
 Gev and are no longer significant at $M \sim 4$ GeV.

Let us note, that our main goal in this model calculation is to
get a relative enhancement of the dilepton rate as compared to
the usual calculations exploiting binary collision approximation.
A well-known drawback of the one-dimensional scaling expansion are the
long expansion times, $\sim 100$ fm already at SPS energies.
 In a more
realistic three-dimensional treatment the relative time scales for
different stages of system`s evolution can change, but the overall
relative enhancement for the contribution from the hadron phase can be
expected to be more stable. In particular, this should be the case
at SPS where the mixed and hadron phases are expected to dominate.

\section{Conclusion}

We have presented a method for taking into account the higher order virial
corrections
to the dilepton emission rate in a hot and dense hadron gas. To
go beyond a binary collision approximation
with a limited number of interacting meson species we argue that the
thermal emission rates of large mass lepton pairs from the quark matter and
dense hadron gas are approximately equal
in the vicinity of $T_c$. We then construct an interpolation between the
hadron rate due to binary collisions at low $T$ and the rate at the critical
point $T=T_c$, effectively the quark matter rate.
The resulting rates are used  in a one-dimensional hydrodynamical
description of heavy ion collisions at   SPS and  RHIC
 energies. The main result in the case of a second order deconfinement
transition is that while at SPS energies the higher
order virial corrections to the dilepton rate from the binary hadron
collisions are
still seen in the total thermal dilepton spectrum, at RHIC energies the
emission from quark matter above $T_c$ dominates in the considered mass range
1.5 GeV $\le M \le$ 4 GeV.
In the case of a first order deconfinement phase transition the rate
enhancement due to new sources is bigger than in the second order case at
SPS energies. At RHIC it can still be seen at the low-mass end of the mass
interval under consideration ($M \sim 2$ GeV) but is no longer important
at $M = 4$ GeV, where the emission from the high temperature QGP phase
becomes dominant.

A. Leonidov acknowledges the warm hospitality at the University of
Jyv\"askyl\"a, where a major part of this work was done.  His work was
partially supported by the Russian Foundation for Basic Research under grant
93-02-3815 and by the Academy of Finland under grant 27574.

\end{document}